# New Coordinates for BTZ Black Hole and Hawking Radiation via Tunnelling

Wenbiao Liu

*Department of Physics, Institute of Theoretical Physics, Beijing Normal University, Beijing 100875, China*

## Abstract

Hawking radiation can usefully be viewed as a semi-classical tunnelling process that originates at the black hole horizon. For the stationary axisymmetric BTZ black hole, a generalized Painleve coordinate system (Painleve-BTZ coordinates) is introduced, and Hawking radiation as tunnelling under the effect of self-gravitation is investigated. The corrected radiation is obtained which is not precise thermal spectrum. The result is consistent with the underlying unitary theory. Moreover, Bekenstein-Hawking entropy of BTZ black hole is not necessarily corrected when we choose appropriate coordinate system to study the tunnelling effect.

## I. INTRODUCTION

The thermal Hawking radiation[1,2] implies the loss of unitary, and then the breakdown of quantum mechanics.[3] According to Hawking radiation, black hole will radiate its energy away, and vanish in the end. Where does the information go? Although Hawking radiation can be described in a unitary theory in string theory,[4] it is not clear how information is returned.

Originally, Hawking explained the existence of black hole radiation as particle's tunnelling coming from vacuum fluctuations near the horizon. The radiation is like electron-positron pair creation in a constant electric field. The energy of a particle can change its sign after crossing the horizon. So a pair created by vacuum fluctuations just inside or outside the horizon can materialize with zero total energy, after one member of the pair has tunnelled to the opposite side. However, Hawking's derivation did not proceed in this way.[1]There were two difficulties to overcome. The first was to find a well-behaved coordinate system at the event horizon. The second was where is the barrier . Recently, a method to describe Hawking radiation as tunnelling process was developed by Kraus and Wilczek[5] and elaborated by Parikh and Wilczek.[6−8] This method gives a leading correction to the emission rate arising from the loss of mass of the black hole corresponding to the energy carried by the radiated quantum. Following this method, the radiation from AdS black hole and de Sitter cosmological horizon were also studied.[9−11] All these spherically symmetric investigations are successful. Because of the complexity of the stationary axisymmetric Kerr black hole,[12] the tunnelling effect must be investigated in the dragging coordinate system. The coordinate system and tunnelling result were successfully given in [13]. The picture is: a particle do tunnel out of Kerr black hole, the barrier is created by the outgoing particle itself. If the total energy and angular momentum must be reserved, the outgoing particle must tunnel out a radial barrier to an observer resting in dragging coordinate system.

The black hole solutions of Banados, Teitelboim and Zanelli[14,15] in (2+1) spacetime dimensions are similar as Kerr black hole due to the axisymmetry, so the dragging coordinate system should be used to investigate tunnelling effect naturally. Because the coordinate system is not appropriate in [16], the entropy expression must be modified in order to get

$$\Gamma = e^{-2\,\mathrm{Im}\,I} = e^{+\Delta S}, \qquad (1)$$

where $\Gamma$ is the emission rate, $I$ is the action for an outgoing positive energy particle, $\Delta S$ is the change in the entropy of black hole. In this paper, we will give a new dragging coordinate system, and show that the complex modified entropy is not necessary in BTZ black hole.

The remainder of the paper is organized as follows. In Sec. 2, BTZ black hole metric will be introduced in details. In Sec. 3, after the dragging coordinate system is given, some features of it will be discussed. In Sec. 4, we calculate the rate of emission, obtain a correction spectrum without modifying the Bekenstein-Hawking entropy expression. In the end, a brief discussion will be given.

## II. BTZ BLACK HOLE

The BTZ black hole[14,15,17−19] is solution of the standard Einstein-Maxwell equation in 2+1 spacetime dimensions, with a negative cosmological constant. Using BTZ black hole metric, we can study black hole in a lower-dimension spacetime which could exhibit the key features without the unnecessary complications. Ignoring the Maxwell field, we can write down the action of a three dimensional theory of gravity as

$$I = \frac{1}{2\pi} \int \sqrt{-g}[R + 2l^{-2}]d^2xdt + B, \qquad (2)$$

where $B$ is a surface term and the radius $l$ is related to the cosmological constant by $-\Lambda = l^{-2}$. The equations of metric derived from Eq.(2) are solved by the black hole field

$$ds^2 = -(-M + \frac{r^2}{l^2} + \frac{J^2}{4r^2})dt^2 + (-M + \frac{r^2}{l^2} + \frac{J^2}{4r^2})^{-1}dr^2 + r^2(d\varphi - \frac{J}{2r^2}dt)^2, \qquad (3)$$

with $M$ the ADM mass, $J$ the angular momentum (spin) of the BTZ black hole and $-\infty < t < +\infty$, $0 \leq r < +\infty$, $0 \leq \varphi \leq 2\pi$.

For the positive mass black hole spectrum with spin ($J \neq 0$), the line element (3) has two horizons:

$$r_\pm^2 = l^2\{\frac{M}{2}[1 \pm (1 - \left(\frac{J}{Ml}\right)^2)^{1/2}]\}. \qquad (4)$$

Of these, $r_+, r_-$ are the black hole outer and inner horizon respectively. In order for the horizon to exist one must have

$$M > 0, \qquad |J| \leq Ml.$$

In the extremal case $|J| = Ml$, both roots of $g_{00} = 0$ coincide.

The area $A_H$ and Hawking temperature $T_H$ of the event (outer) horizon are[20,21]

$$A_H = 2\pi l \{\frac{M}{2}[1 + (1 - \left(\frac{J}{Ml}\right)^2)^{1/2}]\}^{1/2} = 2\pi r_+, \tag{5}$$

$$T_H = \frac{\sqrt{2}}{2\pi l} \frac{\sqrt{M^2 - J^2/l^2}}{(M + \sqrt{M^2 - J^2/l^2})^{1/2}} = \frac{1}{2\pi l^2}(\frac{r_+^2 - r_-^2}{r_+}). \tag{6}$$

The entropy of the spinning BTZ black hole is

$$S = 4\pi r_+, \tag{7}$$

and if we reinstate the Planck units (since in BTZ units $8\hbar G = 1$) we get

$$S = \frac{1}{4\hbar G} A_H = S_{BH},$$

which is the well-known Bekenstein-Hawking area formula ($S_{BH}$) for the entropy, and is proven by counting excited states recently in [22].

### III. PAINLEVE-BTZ COORDINATES

To do a tunnelling computation at the event horizon, we should find a coordinate system that is well-behaved there. At first, we investigate the dragging coordinate system. Let

$$\frac{d\varphi}{dt} = \frac{J}{2r^2} = \Omega, \tag{8}$$

then the line element of BTZ black hole can be rewritten as

$$ds^2 = -(-M + \frac{r^2}{l^2} + \frac{J^2}{4r^2})dt^2 + (-M + \frac{r^2}{l^2} + \frac{J^2}{4r^2})^{-1}dr^2. \tag{9}$$

In fact, the metric Eq.(9) represents a 2-dimensional hypersurface in 3-dimensional BTZ sapcetime. This dragging coordinate system is not what we want to use for resolving tunnelling effect in BTZ spacetime. We need another transformation to make none of the components of either the metric or the contra metric diverge at the horizon. Moreover, constant time slices are just flat Euclidean in radial. To obtain a coordinate system analogous to Painleve coordinates[23], we should perform a coordinate transformation

$$dt = d\tau + f(r)dr, \tag{10}$$

where $f(r)$ is a function of $r$, independent on $t$.

Putting Eq.(10) into the line element expression (9), we have

$$\begin{aligned} ds^2 &= -(-M + \frac{r^2}{l^2} + \frac{J^2}{4r^2})dt^2 + (-M + \frac{r^2}{l^2} + \frac{J^2}{4r^2})^{-1}dr^2 \\ &= -(-M + \frac{r^2}{l^2} + \frac{J^2}{4r^2})d\tau^2 - 2f(r)(-M + \frac{r^2}{l^2} + \frac{J^2}{4r^2})d\tau dr \\ &\quad + [(-M + \frac{r^2}{l^2} + \frac{J^2}{4r^2})^{-1} - (-M + \frac{r^2}{l^2} + \frac{J^2}{4r^2})f^2(r)]dr^2. \end{aligned} \tag{11}$$

As a corollary, we demand that the metric is flat Euclidean in radial to the constant-time slices. We then get the condition

$$(-M + \frac{r^2}{l^2} + \frac{J^2}{4r^2})^{-1} - (-M + \frac{r^2}{l^2} + \frac{J^2}{4r^2})f^2(r) = 1,$$

that is

$$f^2(r) = (-M + \frac{r^2}{l^2} + \frac{J^2}{4r^2})^{-1}[(-M + \frac{r^2}{l^2} + \frac{J^2}{4r^2})^{-1} - 1]. \tag{12}$$

So, the Eq.(11) can be changed into

$$ds^2 = -(-M + \frac{r^2}{l^2} + \frac{J^2}{4r^2})d\tau^2 + 2[-(-M + \frac{r^2}{l^2} + \frac{J^2}{4r^2}) + 1]^{1/2}d\tau dr + dr^2 \tag{13}$$

There is now no singularity at the event horizon, and the true character of the spacetime, as being stationary but not static, is manifest.

For later usage, let us evaluate the radial, null geodesics described by Eq.(13) as following

$$-(-M + \frac{r^2}{l^2} + \frac{J^2}{4r^2}) + 2[-(-M + \frac{r^2}{l^2} + \frac{J^2}{4r^2}) + 1]^{1/2}\dot{r} + \dot{r}^2 = 0,$$

where a dot means derivation with respect to $\tau$. Solving the quadratic, we then have

$$\dot{r} = \frac{dr}{d\tau} = \pm 1 - [1 - (-M + \frac{r^2}{l^2} + \frac{J^2}{4r^2})]^{1/2}, \tag{14}$$

where the +(-) sign can be identified with outgoing (ingoing) radial motion, respectively, under the assumption that $\tau$ increases towards future.

Let us now focus on a semiclassical treatment of the associated radiation. We adopt the picture of a pair of virtual particles spontaneously created just inside the horizon. The positive energy virtual particle can tunnel out while the negative one is absorbed by the black hole resulting in a decrease in the mass. The particle is considered as a shell (an ellipsoid shell) of energy $\omega$. We fix the total ADM mass and let the mass $M$ of the BTZ black hole vary. If a shell of energy (mass) $\omega$ is radiated outwards the outer horizon, the BTZ black hole mass will be reduced to $M - \omega$. We should replace $M$ with $M - \omega$ in the metric (13) and the geodesic Eq. (14) to describe the moving of the shell.

$$ds^2 = -(-(M-\omega) + \frac{r^2}{l^2} + \frac{J^2}{4r^2})d\tau^2 + 2[-(-(M-\omega) + \frac{r^2}{l^2} + \frac{J^2}{4r^2}) + 1]^{1/2} d\tau dr + dr^2, \tag{15}$$

$$\dot{r} = \frac{dr}{d\tau} = \pm 1 - [1 - (-(M-\omega) + \frac{r^2}{l^2} + \frac{J^2}{4r^2})]^{1/2}. \tag{16}$$

### IV. TUNNELLING PROCESS

We evaluate the imaginary part of the action for an outgoing positive energy particle which crosses the horizon outwards from:

$$r_{in}^2 = r_+^2(M, l, J) = l^2 \{ \frac{M}{2}[1 + (1 - \left(\frac{J}{Ml}\right)^2)^{1/2}] \}, \tag{17}$$

to

$$r_{out}^2 = r_+^2(M-\omega, l, J) = l^2 \{ \frac{M-\omega}{2}[1 + (1 - \left(\frac{J}{(M-\omega)l}\right)^2)^{1/2}] \}. \tag{18}$$

The imaginary part of the action is

$$\text{Im} I = \text{Im} \int_{r_{in}}^{r_{out}} p_r dr = \text{Im} \int_{r_{in}}^{r_{out}} \int_0^{p_r} dp_r dr. \tag{19}$$

We make the transition from the momentum variable to the energy variable using Hamilton's equation $\dot{r} = \frac{dH}{dp_r}$. Thinking about Eq.(16) in the vicinity of event horizon, the result is

$$\text{Im}\, I = \text{Im} \int_{r_{in}}^{r_{out}} \int_0^\omega \frac{(-d\omega')dr}{1 - [1 - (-(M-\omega') + \frac{r^2}{l^2} + \frac{J^2}{4r^2})]^{1/2}}, \tag{20}$$

where the minus sign of $d\omega'$ is due to the Hamiltonian being equal to the modified mass $H = M - \omega$. This is not disturbing since $r_{in} > r_{out}$. After some calculations (involving contour integration into the lower half of $\omega'$ plan), we get

$$\text{Im}\, I = 2\pi(r_{in} - r_{out}). \tag{21}$$

Apparently the emission rate depends not only on the mass $M$ and angular momentum (spin) $J$ of BTZ black hole but also on the energy $\omega$ of the emitted massless particle

$$\Gamma(\omega, M, l, J) = e^{-2\,\text{Im}\,I} = \exp[4\pi(r_{out} - r_{in})]. \tag{22}$$

Comparing Eq.(7) and Eq.(22), we have

$$\Gamma(\omega, M, l, J) = \exp(\Delta S), \tag{23}$$

which is just Eq.(1) that we want to get.

## V. CONCLUSIONS AND DISCUSSIONS

After considering the dragging effect in BTZ spacetime, we conclude that the emission rate can be expressed as the standard form Eq.(23). The complex modified entropy in [16] is not necessary, and the entropy of BTZ black hole is just the Bekenstein-Hawking formulism.

Thinking of Eq.(4) and Eq.(6), we have

$$r_{out} - r_{in} = \frac{r_{out}^2 - r_{in}^2}{r_{out} + r_{in}} = \frac{1}{2r_+}l^2[-\frac{\omega}{2} + \frac{M-\omega}{2}(1 - (\frac{J}{(M-\omega)l})^2)^{1/2}$$
$$-\frac{M}{2}(1 - (\frac{J}{Ml})^2)^{1/2}], \tag{24}$$

$$T = \frac{1}{2\pi l^2}(\frac{r_+^2 - r_-^2}{r_+}) = \frac{M}{2\pi r_+}\sqrt{1 - (\frac{J}{Ml})^2}. \tag{25}$$

Using Tailor expansion, we have

$$\sqrt{1-(\frac{J}{(M-\omega)l})^2} = \sqrt{1-(\frac{J}{Ml})^2(1+\frac{2\omega}{M}-(\frac{\omega}{M})^2)}$$
$$= \sqrt{1-(\frac{J}{Ml})^2} - \frac{(\frac{J}{Ml})^2\frac{\omega}{M} - \frac{1}{2}(\frac{J}{Ml})^2(\frac{\omega}{M})^2}{\sqrt{1-(\frac{J}{Ml})^2}}. \tag{26}$$

Putting Eq.(26) into Eq.(24), we get

$$r_{out} - r_{in} = \frac{l^2}{2r_+}\{-\frac{\omega}{2} + \frac{M}{2}(\frac{\frac{1}{2}(\frac{J}{Ml})^2(\frac{\omega}{M})^2}{\sqrt{1-(\frac{J}{Ml})^2}} - \frac{(\frac{J}{Ml})^2\frac{\omega}{M}}{\sqrt{1-(\frac{J}{Ml})^2}})$$
$$-\frac{\omega}{2}[\sqrt{1-(\frac{J}{Ml})^2} - \frac{(\frac{J}{Ml})^2\frac{\omega}{M}}{\sqrt{1-(\frac{J}{Ml})^2}} + \frac{\frac{1}{2}(\frac{J}{Ml})^2(\frac{\omega}{M})^2}{\sqrt{1-(\frac{J}{Ml})^2}}]\}. \tag{27}$$

Thinking of the linear order of $\omega$ in Eq.(27), and taking it into Eq.(22), we have

$$\Gamma(\omega, M, l, J) = \exp(-\beta\omega + C(M, l, J)\omega^2). \tag{28}$$

To linear order of $\omega$, we find that the rate is Boltzmann factor $\exp(-\beta\omega)$ with inverse temperature $\beta$. This is the familiar result. But note that at higher energies the spectrum cannot be approximated as thermal. The precise expression, Eq.(28), can be written as the exponent of the difference in the Bekenstein-Hawking entropy, $\Delta S$, before and after emission[7,24], just as Eq.(23).

Note also that Eq.(28) is consistent with an underlying unitary theory. According to Ref.[6], "Quantum mechanics tells us that the rate must be expressed as

$$\Gamma(i \to f) = |M_{fi}|^2 \cdot (phase\quad space\quad factor), \tag{29}$$

where the first term on the right is the square of the amplitude for the process. The phase space factor is obtained by summing over final states and averaging over initial states. But the number of final states is just the exponent of the final entropy, while the number of initial states is the exponent of the initial entropy." So, we have

$$\Gamma = \frac{e^{S_{final}}}{e^{S_{intial}}} = \exp(\Delta S), \tag{30}$$

which is in agreement with our result Eq.(23). This suggests that the formula we have is actually exact, up to a prefactor.

This research is supported by the National Natural Science Foundation of China (Grant No. 10373003, 10475013) and the National Basic Research Program of China (Grant No. 2003CB716302).